\documentclass[fleqn,twoside]{article}
\usepackage[headings]{espcrc2}

\usepackage{graphicx}
\usepackage[figuresright]{rotating}
\usepackage{amsmath}

\newcommand{\bvec}[1]{\boldsymbol{#1}}
\newcommand{\geant}{\textsc{Geant4}}

\title{Monte Carlo based studies of a polarized positron source for
International Linear Collider (ILC)}

\author{Ralph Dollan\address[hu]{
          Humboldt-Universit{\"a}t zu Berlin, 
          Institut f{\"u}r Physik, Newtonstr.\ 15, 12489 Berlin},
        Karim Laihem\address[zeuthen]{
          DESY Zeuthen, Platanenallee 6, 15738 Zeuthen}, 
        Andreas Sch{\"a}licke\addressmark[zeuthen].
}
       
\runtitle{Studies of a polarized positron source using \geant}

\begin{document}

\begin{abstract}
The full exploitation of the physics potential of an International
Linear Collider (ILC) requires the development of a polarized positron
beam. New  concepts of 
polarized positron sources are based on the development of circularly
polarized photon sources. The polarized photons create
electron-positron pairs in a thin target and transfer their
polarization state to the outgoing leptons. To achieve a high level of
positron polarization the understanding of the production mechanisms
in the target is crucial.
Therefore a general framework for the simulation of polarized
processes with \geant\ is under development. In this contribution the
current status of the project and its application to a 
study of the positron production process 
for the ILC is presented.  
\vspace{1pc}
\end{abstract}

\maketitle

\section{Introduction}

A future International Linear Collider (ILC) provides an outstanding
tool for the precise exploration of physics at TeV scale
\cite{ILC}. In contrast to hadron colliders, the well defined initial
state and the cleanliness of the final states allow for a precise
measurement of Standard Model and new physics processes. 
Having both, positron and electron, beams polarized will be a decisive
improvement for physics studies. 
A recent review of the physics case of ILC using polarized positrons
can be found in \cite{Moortgat-Pick:2005cw}.

\begin{figure}[htb]
\vspace*{-2mm}
\includegraphics[width=78mm]{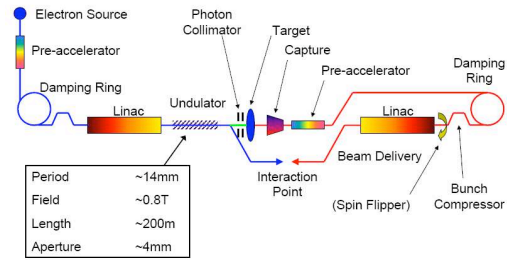}
\vspace*{-12mm}
\caption{Possible layout of an ILC. Polarized positrons are created
from polarized photons produced in a {\em helical undulator}.}
\label{fig:ilclayout}
\end{figure}

One possible layout for the production of polarized positrons is
sketched in figure \ref{fig:ilclayout}. Circularly polarized photons
are created by  sending the electron beam through a
{\em helical undulator} \cite{Mikhailichenko}. In a thin target the 
photons are converted into polarized positrons via pair creation. 
In comparison  to a conventional positron source this method
substantially reduces the heat load in the positron target. 

A demonstration experiment to quantify the yield of polarization of an 
undulator based positron source, 
E166 \cite{Alexander:2003fh}, is currently performed at SLAC. 
The experience gained with E166 is decisive for designing and optimising the 
polarized positron source of the ILC. The precise simulation of the
positron production as well as the polarimetry processes is essential
for a complete understanding of the data obtained at E166.
A simulation toolkit for the investigation of polarized
processes based on \geant\ is currently under development. 
In this contribution a status report of this project is given.

\section{Implementing polarization into \geant}

\geant\ is a toolkit for the simulation of the passage of particles
trough matter \cite{Agostinelli:2002hh}. Individual particles are
tracked step by step and each step can lead to creation of particles,
destruction of particles, or to a modification of the particle 
properties.  
The aspect of polarization has so far been widely neglected
\footnote{The only polarized process supported by the
current release of \geant\ is Compton scattering of linear polarized,
low-energy photons on an unpolarized target.}. 
With our extension it will be possible to track also polarized
particles (leptons and photons). Special emphasis will be put in the
proper treatment of polarized matter, which is essential for the
simulation of positron polarimetry. 
It is planned to create a universal framework for polarization
and to implement it in an official \geant\ release.

To realise this project, the following polarization
dependent processes have to be considered
\begin{itemize}
\item Compton scattering,\\[-6mm]
\item Bhabha/M{\o}ller scattering,\\[-6mm]
\item Pair creation,\\[-6mm] 
\item Bremsstrahlung.
\end{itemize}
In addition to these well localised interactions, the influence of 
magnetic fields on the electron (or positron) spin has to be treated
properly. 

In the following section, a brief review of existing simulation tools
for polarization transfer is given. In the subsequent sections the
proposed framework for \geant\ is presented. 

\subsection{Existing codes for the simulation of polarized processes}

Several simulation packages for the realistic description
of the development of electromagnetic showers in matter have been
developed. A prominent example of such codes is EGS (Electron Gamma 
Shower)\cite{Nelson:1985ec}. 
For this simulation framework extensions with the treatment of
polarized particles exist \cite{Floettmann:thesis,Namito:1993sv,Liu:2000ey}; 
the most complete 
has been developed by K.~Fl{\"o}ttmann \cite{Floettmann:thesis}. It is
based on the matrix formalism \cite{McMaster:1961}, which enables a
very general treatment of polarization. However, the Fl{\"o}ttmann
extension concentrates on evaluation of polarization transfer, i.e.\
the effects of polarization induced asymmetries are neglected, and
interactions with polarized media are not considered. 

Another important simulation tool for detector studies is \textsc{Geant3}
\cite{Brun:1985ps}. Here also some effort has been made to include
polarization \cite{Alexander:2003fh,Hoogduin:thesis}, but these
extensions are not publicly available.

\subsection{Polarization framework for \geant}

The package \geant\ is the newest member on the simulation
front. It is entirely written in C++. It has a wide range of
application, and slowly replaces the Fortran based simulation toolkits.

The proposed implementation of polarized processes is based on Stokes
vectors and allows a convenient description of the polarization
transfer by the matrix formalism \cite{McMaster:1961}.
In this formalism, a three-component {\em polarization vector}
$\bvec{\xi}$ is assigned  to each particle and characterises
completely  the polarization state of any lepton or
photon\footnote{This vector is already present in the current release 
of \geant, but it is only used in low-energy Compton scattering of
linear polarized photons. The interpretation as Stokes vector allows
for the usage in a more general framework.}. 
For the simulation of polarized media, a possibility to assign Stokes
vectors to physical volumes has to be provided in \geant. This 
is handled by a new class, the so-called {\em polarization manager}. 
It also allows the evaluation of Stokes vectors in different frames of
reference.  

The general procedure is very similar to the polarization extension to
EGS by Fl{\"o}ttmann \cite{Floettmann:thesis}. Any interaction is
described by a transfer matrix $T$, which characterises the process
completely. It usually depends on kinematic variables like energy and
angle, but it can also depend on polarization states (e.g. of the
media). The final state polarization $\bvec{\xi}$ is determined via
matrix multiplication with the incoming Stokes vector $\bvec{\xi}_0$,
\begin{align}
  \Biggl(
 \begin{array}{c} 
    I \\
 \bvec{\xi}   
 \end{array}
  \Biggr) = T \,
  \Biggl(
 \begin{array}{c} 
    I_0 \\
 \bvec{\xi}_0   
 \end{array}
  \Biggr)
 \;.
\end{align}
The components $I_0$ and $I$ refer to the incoming and outgoing
intensities, respectively. In this framework the transfer matrix $T$  
is of the form
\begin{align}
 T =
  \begin{pmatrix}
     S   &   A_1    &  A_2    &  A_3    \\
     P_1 &   M_{11} &  M_{21} &  M_{31} \\
     P_2 &   M_{12} &  M_{22} &  M_{32} \\
     P_3 &   M_{13} &  M_{23} &  M_{33} \\
  \end{pmatrix}
 \;.
\end{align}
The matrix elements $T_{ij}$ can be identified as (unpolarized)
differential cross section ($S$), polarized differential cross section
($A_j$), polarization transfer ($M_{ij}$), and (de)polarization ($P_i$).
In the Fl{\"o}ttmann extension the elements $A_j$ and $P_i$ have been
neglected, thus concentrating on polarization transfer only. 
Using the full matrix takes now all polarization effects into account.
The structure is illustrated with a few examples in the following
section. 

\section{Applications}

Here, some preliminary results shall illustrate the field of 
application.

\subsection{Polarized Compton scattering}
\begin{figure}[htb]
\vspace*{-8mm}
\includegraphics[width=78mm]{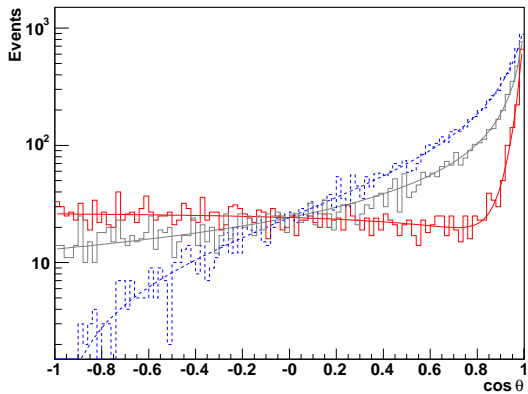}
\vspace*{-10mm}
\caption{Comparison of the \geant\ implementation (histogram) of
polarized Compton scattering with an analytic formula (solid
lines). The graph shows the dependence of the scattering angle on the
polarization states of target (electron) and beam (photon).}
\label{fig:compton}
\end{figure}
The first process studied is Compton scattering. This process
possesses all basic features: a polarization dependent differential
and total cross section, polarization transfer and depolarization
effects. Compton scattering is of great importance in polarimetry. 

In figure \ref{fig:compton} the angular distribution of the scattered
photon is presented.  
For this simulation a 100\% circularly polarized photon beam and a 100\%
longitudinally polarized iron foil is assumed\footnote{Note,
that this is an academic case, since the maximal degree of
polarization in iron is $2/26\approx 7.69$\%.}. When flipping the
electron spin from an anti-parallel configuration 
with respect to the photon spin (blue) to a parallel orientation  
(red), the distribution changes drastically, and 
the total cross section decreases. For illustration,
the case where both, target and beam, are unpolarized is also plotted
(black). A comparison with an analytic formula (solid lines) shows
perfect agreement in all cases.

In a next step a more realistic simulation of target properties will
be performed to study the effects of different polarized processes. 

\subsection{ILC positron source studies}

The polarization transfer in a undulator based positron source
has been investigated. 
Since the target is unpolarized, the total cross section, i.e.\ the
interaction length, does not depend on the photon
polarization. Consequently it is sufficient to concentrate on
the polarization transfer from incoming photons to outgoing positrons. 

\begin{figure}[bt]
\vspace*{-2mm}
\includegraphics[width=70mm]{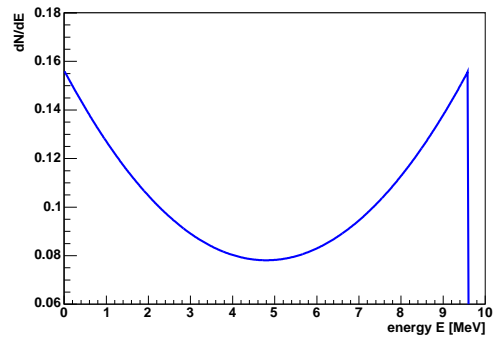}
\vspace*{-10mm}
\caption{First harmonic of the photon energy distribution as created
by a {\em helical undulator} \cite{Mikhailichenko}, for electron
energy $E_e=$50 GeV, undulator period $\lambda_u=$2.4 mm and undulator
strength parameter $K=$0.17. The peak of the first harmonic (dipole)
radiation is at 9.62 MeV.} 
\label{fig:gammaenergy}
\end{figure}
\begin{figure}[t]
\includegraphics[width=78mm]{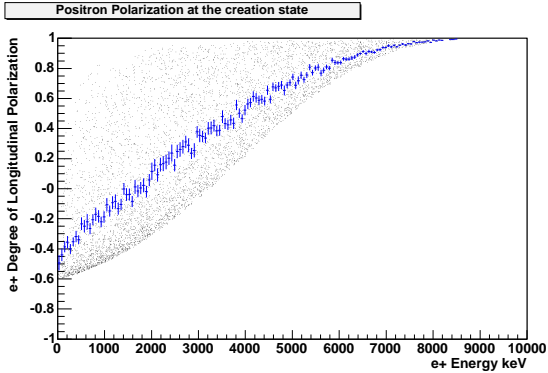}
\vspace*{-10mm}
\caption{Degree of polarization of created positrons. The degree of
polarization of the incoming photon beam is fixed to 100\%, 
the energy spectrum of the photons is given in fig.\
\ref{fig:gammaenergy}. Each dot corresponds to a single answer of the
transfer matrix. The mean degree of polarization is plotted as the
(blue) profile histogram.} 
\label{fig:posipol1}
\end{figure}

The setup of the simulation consists of an incoming photon beam with
the characteristic energy spectrum of a {\em helical undulator}, cf.\ fig.\
\ref{fig:gammaenergy}. As a first approximation, the polarization  of
the photon beam is assumed to be 100\%. 
The degree of polarization of the positrons created in pair 
production depends also on the energies of the incoming photon and the
outgoing positron. In general the degree of the positron polarization
is increasing with the energy fraction of the created positron, see
figure \ref{fig:posipol1}. 

\begin{figure}[htb]
\includegraphics[width=78mm]{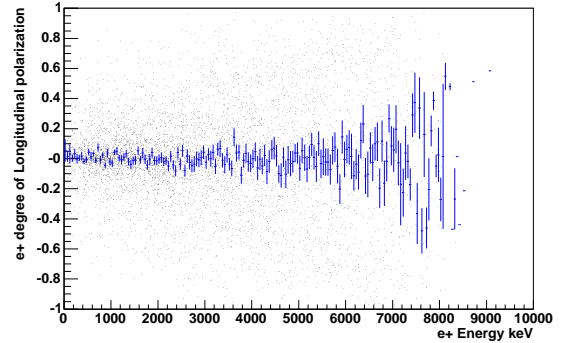}
\vspace*{-10mm}
\caption{Check of the polarization routine: Degree of polarization of
positrons created from 
randomly polarized photons. }  
\label{fig:posipol2}
\end{figure}

A simple first check of the polarization routine is provided by
assuming an equal mixture of left and right circularly polarized photons
as incoming beam. In this case one expects to obtain 
an unpolarized
positron beam. Indeed, in figure \ref{fig:posipol2} this behaviour can be
observed. 

\begin{figure}[htb]
\vspace*{-8mm}
\includegraphics[width=70mm]{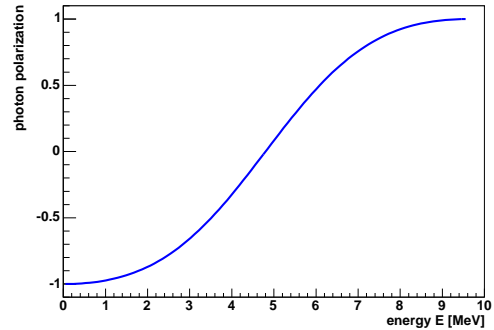}
\vspace*{-10mm}
\caption{Degree of polarization of photons produced in a 
  {\em helical undulator}.} 
\label{fig:gammapol}
\end{figure}

\begin{figure}[htb]
\centerline{\includegraphics[width=73mm]{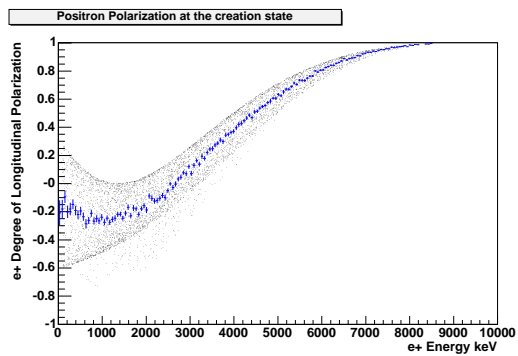}}
\vspace*{-10mm}
\caption{Degree of positron polarization created by photons produced
in a {\em helical undulator} using a realistic spectrum of photon energy
and photon polarization. } 
\label{fig:posipol3}
\end{figure}

Now the polarization spectrum of the 
{\em helical undulator} as 
plotted in figure \ref{fig:gammapol} will be included in the study.
The  simulation shows  the  marginal influence on the obtained degree
of positron polarization. In particular, high energy positrons are
nearly 100\% polarized, see figure \ref{fig:posipol3}.

\begin{figure}[htb]
\includegraphics[width=70mm]{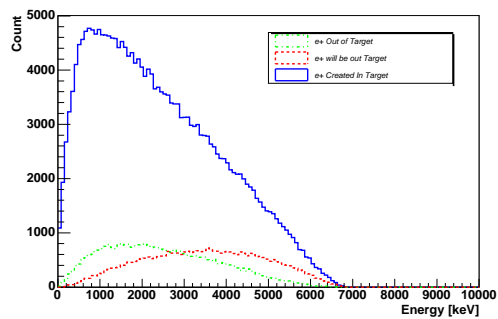}
\vspace*{-10mm}
\caption{Positron energy distribution. The energy of all produced
positrons at the creation point (blue) is compared with the energy of
the positron fraction that will eventually manage to leave the target
(red), and the energy of these positrons at the exit point of the
target (green).} 
\label{fig:posienergy}
\end{figure}

For a realistic simulation of a polarized positron source based on a
{\em helical undulator} the effects of bremsstrahlung, multiple
scattering, Coulomb and screening correction
have to be taken into account. In figure
\ref{fig:posienergy} the influence of these processes on the obtained
positron energy spectrum is investigated. 
It is shown, that only a small fraction (green) of all produced
positrons (blue histogram) will escape from the target. 
The energy spectrum of positrons that leave the target is shifted to
lower values in comparison to their spectrum at the creation point
(red). Consequently, the target acts as a filter for high energy
positrons and the created positrons have suffered a substantial loss
of energy. The effect of bremsstrahlung and multiple scattering on the
degree of polarization of the produced positrons will be the subject
of further investigations. 

\section{Conclusion}
In this report the current status of a project to implement
polarization into the framework of \geant\ has been presented. For this
task the Stokes formalism is employed, providing a systematic approach
for a consistent treatment of polarized leptons and photons.
Some preliminary results demonstrate the applicability of this new
extension to polarimetry (Compton scattering) and polarization
transfer studies (positron source). These analyses represent the first
steps toward a realistic target simulation of an undulator based
positron source for the ILC.

\section*{Acknowledgement}
The authors are indebted to A.\ Stahl as the initiator of this project,
and also like to thank T.~Lohse for fruitful collaboration, and
S.~Riemann, P.~Starovoitov and J.~Suarez 
for helpful discussions. 
K.L. and A.S. are grateful for the assistance by K.\ Fl{\"o}ttmann and
J.~C.~Liu concerning the EGS polarization extension.
R.D.\ acknowledges support by the European Community (EUROTeV contract
number RIDS~011899).

\end{document}